# Knowing when to stop

Ofer Lahav and Joseph Silk

**When do we stop an ongoing science project to make room for something new? Decision-making is a complex process, ranging from budgetary considerations and tension between ongoing projects, to progress assessment and allowance for novel science developments.**

We make numerous decisions in everyday life on when to conclude or stop a process. Some are easier than others. When to cross the road? How to choose a spouse? How many candidates to short list for a job? And the focus of this Comment: when do we terminate costly science projects, notably in cosmology to open windows to new research directions?

In the world of science, we have to decide on what problems to tackle, which techniques and instruments to use, but also when to stop collecting and analysing data if resolution of a specific question becomes too compromised or too elusive, or simply too expensive. Of course, the same data set can be used to look at entirely new questions or may yield unexpected discoveries. What is sufficient? How to optimally choose when driven by the combination of science return and budgetary limits?

## 1. The current cosmological model

While the LCDM (the Cosmological Constant Lambda + Cold Dark Matter) model has passed many observational tests (e.g. [1] for review), it is disturbing that we still don't know the nature of its two dark components, dark matter and dark energy. There are also 'tensions' in certain parameters of LCDM when measured by different probes. In particular, there are concerns around a tension of as much as 4 standard deviations (sigma) in the value of the Hubble constant $H_0$ when contrasting the remote, cosmic microwave background (CMB), results ([2]) with local, supernova and stellar-based, measurements ([3]; [4]), and about 2-sigma in the matter 'clumpiness' parameter $S_8$. The latter derives from weak gravitational lensing surveys ([5] ;[6] ; [7]) that indicate a smoother mass distribution than that measured from the CMB by the ESA'S Planck space mission. While usually one may not get too concerned about 'just 2-sigma', should we search for 'new Physics' if three independent experiments indicate similar 2-sigma disparity?

The community of observational cosmology has been very fortunate over the past two decades to receive funding for many ambitious projects across the world. These projects have already produced rich data sets, broadly supporting the LCDM model. Projects under construction or at the start of operations will yield samples of billions of galaxies at the cost of billions of dollars. The cost of CMB searches is some 100 times more per sky pixel or information bit, but uniquely probes the very early Universe.

## 2. How far to go?

Should we optimize the science return of projects and how do we prioritize them? What are the most important parameters of interest? How accurately (finding the correct values) and precisely (minimizing the uncertainty) do we need to measure these? If different experiments provide different answers, how do we evaluate the cumulative significance of these tensions?

We suggest that there are three categories of parameters, based on the reliability of current measurements. In the first category the four density parameters, the Omega's, (baryons, cold dark matter, dark energy and curvature) are known reasonably well. We remark that while new physics could be hiding in the systematic uncertainties, we may have reached the point of diminishing returns for our science investment. In the second category, additional effort in coming years is required to converge the Dark Energy equation of state *w.* Any hints of surprises are meager; our future efforts focus on convergence.

We place in the third category the Hubble Constant $H_0$, the clumpiness factor $S_8$, the neutrino masses and the CMB polarization tensor-to-scalar ratio, where it is worth putting significant effort in obtaining additional data and innovative techniques towards pinning down these parameters. Several potential sources of systematics still need to be overcome to arrive, for example, at percent precision in $H_0$ determinations and in delensing the CMB foregrounds. Here there is a subtle difference: we know that neutrinos possess mass, but there is no guarantee of a robust inflationary signal of the imprint of gravity waves. Theory suggests that new physics could be hiding in the systematics, and unveiled as our limits improve.

We note that certain cosmological analyses, for example, galaxy surveys and the CMB, involve a package of cosmological parameters that are all derived simultaneously from Bayesian pipelines. But there is freedom to choose priors on parameters that we already know well, and there is flexibility to add more external constraints, such as priors on neutrino mass from terrestrial experiments, to improve constraints on neutrino mass.

We should also be aware of `confirmation bias' — that is, whether the value measured is the value that people 'like', for reasons of simplicity, perceived beauty or deep underlying theory, as exemplified by zero curvature of the Universe or, $w = -1$ (equivalent to the cosmological constant). In contrast, there is nothing fundamental about the value of the $H_0$, apart from requiring its value to satisfy a range of other observations, for example the age of the Universe. To overcome concerns about confirmation biases, it is common now to perform `blinding' in cosmology analyses. But the `devil is in the detail' on how that blinding is implemented, and what to do if the 'unblinding' result is not the `favourite' answer.

The precision of the measurement matters of course, especially with respect to theories. For example, observations of the Dark Energy Survey (DES) weak lensing and clustering combined with CMB's Planck and other probes currently gives $w = -1.03 \pm 0.03$ ([8]). If two competing theories differ by only a small amount it would still be interesting to conduct more measurements rather than to rely on the prejudice that $w=-1$ exactly.

But where do we draw the line? Should we wait for experimental convergence of these theories? Pending such an eventuality, how do we rank the array of alternative theories, especially given that current support for theory is increasingly an afterthought in major experiments? One example, modified gravity, is largely motivated by our inability to identify dark matter and dark energy. We have only explored an infinitesimal part of the dark matter parameter space. Should we prioritize the least costly dark matter searches selected to potentially boost science return per dollar? How do we evaluate the various motivations that underlie targeted dark matter searches? And when do we stop?

Theory poses similar challenges. How complex a dark matter theory should we consider? Should simplicity (Occam's razor arguments), elegance or even beauty play a role in our landscape of alternative theories? And is verifiability even necessary?

## 3. Exploring the unknowns

Fortunately, even if surveys were designed for Dark Energy probes or experiments built for Dark Matter detection, more than just cosmological information can be extracted. Serendipity has proved to be a major element in scientific progress. It is challenging to compare a search that focuses on serendipitous exploration of a greatly expanded detection space with one that targets precision of known elements. Current strategies limit funds for discovery space exploration and we need to carefully manage our resources for optimal science extraction.

Exploration of virgin territory is a guaranteed way of amplifying science return. The dark ages, before the first stars, is the latest horizon for cosmology. Were we able to explore the Universe at a redshift of up to 100, we would potentially unveil unknown phenomena but above all, allow exploration of our cosmic origins to an unprecedented extent. The cost of a definitive experiment is potentially huge. Should this be done at the expense of, for example, delaying implementation of comparably costly exploration of another largely unexplored horizon such as CMB spectral distortions arising from the first years of the Universe?

## 4. Looking forward

Limited resources may compel us to pick the low hanging fruit, even if the dietary content is meager, to better design our telescopes of the future. Budgetary issues mean that use of our finite resources must be prioritized. A layered approach involving different timelines from near to far future is required. The planned projects are important, but we advocate continuously revising their scientific prioritisation based on knowledge derived from existing projects. And above all, we must look further ahead to anticipate unexplored directions as different windows of opportunity arise.

We are used to shutting down smaller projects to make room for future developments. Or more commonly, transferring them to private resources.
Our philosophy is not original. Its application is, however, as we focus on the motivating science goals. We need to drop some to make room for others.

How far ahead? The planned 100 TeV collider will most likely not see action until 2050. ESA is prioritizing its science missions over 2035-2050. To build the strongest case for cosmology, we need to focus on key future goals that are inaccessible to currently planned projects. To access the funding wedge to the future, we need to be innovative. We need to reassess our priorities. The next real advance in cosmology may be via low frequency radio telescopes on the far side of the Moon or with future telescopes of unprecedented scale. To acknowledge our demands of the projects and plant the seeds and initiate the resources for the next generation of astronomers, we must reduce the scale of some existing windows by acknowledging that no substantial gains are likely to be obtained, supported by developing optimal strategies (e.g. [9]). Awareness and assessment of the changing landscape can help the entire community.

*Ofer Lahav is at the Department of Physics & Astronomy, University College London, Gower Street, London, WC1E 6BT, United Kingdom. Joe Silk is at the Institut d'Astrophysique de Paris (IAP), 98 bis Bd Arago, F-75014 Paris, France.*
*e-mail: [o.lahav@ucl.ac.uk](o.lahav@ucl.ac.uk), [silk@iap.fr](silk@iap.fr)*

**References:**
[1] Peebles, P.J.E, arXiv:2106.02672 (2021)
[2] Planck collaboration, A&A, 641, A6 (2020)
[3] Riess, A.G. et al. Ap J Lett, 908, L6 (2021)
[4] Freedman, W., arXiv:2106.15656, ApJ (2021), in press.
[5] Hikage, C., et al. (HSC team), PASJ, 71, 43 (2019)
[6] Asgari, M., et al. Astro. Astrophys. 645, A104 (2021).
[7] Secco, L., et al. (DES collaboration), arXiv:2105.13544 (2021)
[8] DES collaboration, arXiv:2105.135490 (2021)
[9] Trotta, R., Contemporary Physics, 49, 71 (2008)